

\def\svec#1{\skew{-2}\vec#1}
\def\tr{\,{\rm tr}\,}
\magnification=1200
\hoffset=-.1in
\voffset=-.2in

\vsize=7.5in
\hsize=5.6in
\tolerance 10000

\baselineskip 12pt plus 1pt minus 1pt
\smallskip
\centerline{\bf CHERN--SIMONS SOLITONS, TODA}
\centerline{{\bf THEORIES AND THE CHIRAL
MODEL}\footnote{*}{This work is supported in part by funds
provided by the U. S. Department of Energy (D.O.E.) under contract
\#DE-AC02-76ER03069, and NSF grant \#87--08447.}}
\vskip 24pt
\centerline{ Gerald Dunne}
\vskip 12pt
\centerline{\it Department of Mathematics}
\vskip 12pt
\centerline{and}
\vskip 12pt
\centerline{\it Center for Theoretical Physics}
\centerline{\it Massachusetts Institute of Technology}
\centerline{\it Cambridge, Massachusetts\ \ 02139\ \ \ U.S.A.}
\vskip 1.5in
\centerline{Submitted to: {\it Communications in Mathematical Physics\/}}
\vskip .5in
\centerline{Suggested running title: ``Chern--Simons Solitons}
\vfill
\centerline{ Typeset in $\TeX$ by Roger L. Gilson}
\vskip -12pt
\noindent CTP\#2079\hfill March 1992
\eject
\baselineskip 24pt plus 2pt minus 2pt
\centerline{\bf ABSTRACT}
\medskip
The two-dimensional self-dual Chern--Simons equations are equivalent to the
conditions for static, zero-energy solutions of the $(2+1)$-dimensional gauged
nonlinear Schr\"odinger equation with Chern--Simons matter-gauge dynamics.
In this paper
we classify all finite charge $SU(N)$ solutions by first transforming the
self-dual Chern--Simons equations into the two-dimensional chiral model (or
harmonic map) equations, and then using the Uhlenbeck--Wood classification of
harmonic maps into the unitary groups.  This construction also leads to a new
relationship between the $SU(N)$ Toda and $SU(N)$ chiral model solutions.
\vfill
\eject
\noindent{\bf 1.\quad INTRODUCTION}
\medskip
The study of the nonlinear Schr\"odinger equation in $2+1$-dimensional
space-time is partly motivated by the well-known successes of the
$1+1$-dimensional nonlinear Schr\"odinger equation.  Here we consider a {\it
gauged\/} nonlinear Schr\"odinger equation in which we have not only the
nonlinear potential term for the matter fields, but also we have a coupling of
the matter fields to gauge fields.  Furthermore, this matter-gauge  dynamics is
chosen to be of the Chern--Simons form rather than the conventional
Yang--Mills form.  Such a choice is motivated by the fact that the resulting
Schr\"odinger equation is related to a non-relativistic field theory for the
many-body anyon system.

The theory with an Abelian gauge field was analyzed by Jackiw and Pi [9] who
found static, zero energy solutions which arise from a two-dimensional notion
of self-duality.  The static, self-dual matter density satisfies the Liouville
equation which is known to be integrable and, indeed, solvable in the sense
that the general (real) solution may be expressed in terms of an arbitrary
holomorphic function [13].  The gauged nonlinear Schr\"odinger equation with
a {\it non-Abelian\/} Chern--Simons matter-gauge dynamics has also been
considered [6,3,4], and once again static, zero energy solutions (referred
to as self-dual Chern--Simons solitons) have been found to arise from an
analogous, but much richer,  two-dimensional self-duality condition.  These
two-dimensional self-duality equations are formally integrable and as special
cases they reduce to the classical and affine Toda equations, both well-known
integrable nonlinear systems of partial differential
equations [18,15,7,14,1].   For the classical Toda equations one can
exhibit explicit exact solutions with finite charge [6,3,4] using the
results of Kostant [11] and Leznov and Saveliev [12] concerning the
integration of the Toda equations.

In this paper I classify {\it all\/} finite charge solutions to the self-dual
Chern--Simons equations.  This classification is achieved by first showing
that the self-duality equations are equivalent to the classical equations of
motion of the (Euclidean) two-dimensional chiral model (also known as the
harmonic map equations).  One can then make use of some deep classification
theorems due to K.~Uhlenbeck [19] (see also subsequent work by
J.~C.~Wood [23]) which classify all chiral model solutions (for $U(N)$ and
$SU(N)$) with finite classical chiral model action.  The chiral model action
is in fact proportional to the net gauge invariant {\it charge\/} $Q$ in the
matter-Chern--Simons system, and so the classification of all finite charge
solutions is complete.

Another harmonic map result, due to G.~Valli [20], shows that the charge $Q$
is actually {\it quantized\/}, in integral multiples of $2\pi\kappa$ (where
$\kappa$ is the Chern--Simons coupling strength) --- a fact already observed
for the special case of the classical Toda solutions found in Ref.~[3].

The explicit description of $SU(N)$ chiral model solutions rapidly becomes
algebraically complicated, even for $N\ge 3$.  Wood [23] has given an
explicit parametrization involving sequences of holomorphic maps into
Grassmannians and an algorithm involving only algebraic and integral transform
operations.  The $SU(3)$ and $SU(4)$ cases have been studied in great detail
in Refs.~[17,23].  In this paper I present the explicit ``uniton''
decomposition of a class of solutions to the $SU(N)$ chiral model equations
for {\it any\/} $N$.  The matrices are expressed in terms of $(N-1)$ arbitrary
holomorphic functions, and this class of solutions has the remarkable property
that when the matter density matrix is diagonalized it satisfies the classical
$SU(N)$ Toda equations.  At first sight, such a direct correspondence between
the Toda equations and the chiral model equations seems very surprising, but
within the context of the self-duality equations the correspondence arises
quite naturally.

The outline of this paper is as follows.  In Section~2, I describe the
derivation of the self-dual Chern--Simons equations as the conditions (in
fact, necessary and sufficient) for the
static, zero energy solutions to the $2+1$-dimensional gauged nonlinear
Schr\"odinger equation with Chern--Simons coupling.  In Section~3, I show how
the self-duality equations reduce to the integrable Toda equations in special
cases.  The equivalence between the self-dual Chern--Simons equations and the
chiral model (or harmonic map) equations is demonstrated in Section~4, and in
Section~5 I show how to classify all solutions using the results of Uhlenbeck
and Wood.  Finally, in Section~6 I present a special class of explicit
harmonic maps which correspond (via a unitary transformation) to the known
classical $SU(N)$ Toda solutions of the self-dual Chern--Simons equations.  A
brief conclusion is devoted to comments and suggestions for further
investigation.
\goodbreak
\bigskip
\noindent{\bf 2.\quad THE SELF-DUAL CHERN--SIMONS EQUATIONS}
\medskip
The gauged non-linear Schr\"odinger equation reads
$$iD_0\Psi = - {1\over 2} {\svec D}^2 \Psi + {1\over\kappa} \left[
\Psi^\dagger,
\Psi\right] \Psi \eqno(1)$$
where the gauge covariant derivative is $D_\mu \equiv \partial_\mu + \left[
A_\mu,~~~\right]$, and both the gauge potential $A_\mu$ and the matter field
matrix $\Psi$ are Lie algebra valued: $A_\mu = A^a_\mu T^a$, $\Psi = \Psi^a
T^a$.  The Lie algebra generators satisfy:
$$\left[ T^a, T^b\right] = f^{abc} T^c\ \ ,\qquad T^{a\dagger} = - T^a\ \
,\qquad \tr \left( T^a T^b \right) = - {1\over 2} \delta^{ab}\ \ .\eqno(2)$$
Our results are presented for the Lie algebra of $SU(N)$, but the formulation
generalizes naturally for any compact Lie algebra.  In $2+1$ dimensions we
choose to couple the matter and gauge fields via the Chern--Simons equation
$$F_{\mu\nu} = {1\over\kappa} \epsilon_{\mu\nu\rho} J^\rho \eqno(3)$$
where $F_{\mu\nu} = \partial_\mu A_\nu - \partial_\nu A_\mu + \left[ A_\mu,
A_\nu\right]$ is the gauge curvature, $\kappa$ is a coupling constant and
$J^\rho$ is the covariantly conserved $\left( D_\mu J^\mu = 0 \right)$ matter
current
$$\eqalign{ J^0 &= \left[ \Psi,\Psi^\dagger\right] \cr
J^i &= {1\over 2} \left( \bigl[ \Psi^\dagger, D_i \Psi\bigr] - \bigl[ \left(
D_i\Psi\right)^\dagger,\Psi\bigr]\right) \ \ .\cr}\eqno(4)$$
We can also define the scalar current $V^\mu$,
$$\eqalign{V^0 &= \tr\left( \Psi^\dagger\Psi\right) \cr
V^i &= {1\over 2} \tr \left( \Psi^\dagger D_i \Psi - \left( D_i
\Psi\right)^\dagger \Psi\right) \ \ ,\cr}\eqno(5)$$
which is ordinarily conserved $\left( \partial_\mu V^\mu = 0 \right)$.

Note that the Schr\"odinger equation (1) and the Chern--Simons equation (3)
are invariant under the gauge transformation
$$\eqalign{\Psi &\longrightarrow g^{-1} \Psi g \cr
A_\mu &\longrightarrow g^{-1} A_\mu g + g^{-1} \partial_\mu g\ \
,\cr}\eqno(6)$$
where $g\in SU(N)$.  Furthermore, the Schr\"odinger equation (1) may be
expressed as the Heisenberg equation of motion
$$i\partial_0 \Psi = {\delta H\over \delta\Psi^\dagger} \eqno(7)$$
where the Hamiltonian $H$ is
$$\eqalign{
H &= - \int d^2x\,\tr \left( \left( D_+ \Psi^\dagger\right) D_- \Psi\right)\cr
&= {1\over 2} \int d^2x\left( D_+ \Psi^\dagger\right)^a \left(
D_-\Psi\right)^a \ \ .\cr}\eqno(8)$$
Here $D_\pm\equiv D_1\pm i D_2$ and the gauge fields $A_\pm$ appearing in
$D_\pm$ are determined by (3) -- (4).
Note that the simple, manifestly positive,
form (8) for the Hamiltonian relies on the fact that we have chosen the
nonlinear coupling coefficient in (1) to be $1/\kappa$, the {\it same\/} as the
Chern--Simons coupling strength in (3).

We begin by seeking solutions which satisfy the {\it self-dual Ans\"atz\/}
$$D_-\Psi = 0\ \ .\eqno(9)$$
{}From (8) we see that such solutions minimize the energy and thus, by (7),
must
correspond to {\it static\/} solutions.  This fact, together with the
self-dual {\it  Ans\"atz\/} (9), leads to the following concise form of the
Chern--Simons equation (3):
$$\partial_- A_+ - \partial_+ A_- + \left[ A_-,A_+\right] = {2\over\kappa}
\left[ \Psi^\dagger,\Psi\right]\ \ .\eqno(10)$$
Here $A_\pm = A_1 \pm i A_2$, $\partial_\pm=\partial_1 \pm i \partial_2$, and
$A_\pm = - \left( A_\pm\right)^\dagger$.  Equations (9) and (10) are
referred to collectively as the {\it self-dual Chern--Simons equations\/}.
{}From the above discussion we see that solutions of these equations yield {\it
static, minimum\/} (in fact, zero) energy solutions to the gauged nonlinear
Schr\"odinger equation (1) with Chern--Simons coupling (3).  In fact, owing to
a remarkable dynamical $SO(2,1)$ symmetry of (1) and (3), it is possible to
show that {\it all\/} static solutions of (1) and (3) must be self-dual [3].

To conclude this summary we recall that the self-dual Chern--Simons equations
(9) and (10) have arisen previously in another context [8,2]
 --- they are the
dimensional reduction (from Euclidean four dimensions to Euclidean two
dimensions) of the four-dimensional self-dual Yang--Mills equations [6,3,4]
\goodbreak
\bigskip
\hangindent=21pt\hangafter=1
\noindent{\bf 3.\quad ALGEBRAIC REDUCTION TO CLASSICAL AND AFFINE TODA
\hfill\break EQUATIONS}
\medskip
\nobreak
Before classifying the general solutions to the self-dual Chern--Simons
equations we show that certain simplifying algebraic {\it Ans\"atze\/} for the
fields reduce (9) and (10) to familiar integrable nonlinear equations.  First,
choose
$$\eqalignno{A_i &= \sum\limits_\alpha A^\alpha_i H_\alpha &(11\hbox{a}) \cr
\Psi &= \sum\limits_\alpha \psi^\alpha E_\alpha\ \ ,&(11\hbox{b}) \cr}$$
where the sums are over all positive, simple roots $\alpha$ of the algebra
(for $SU(N)$ we may take $\alpha=1\ldots N-1$), and $H_\alpha$, $E_\beta$ are
the Cartan subalgebra and step operator generators (respectively) in the
Chevalley basis [10].
In this Lie algebra basis, the commutation relations have an
especially concise form (for $\alpha,\beta$ simple roots):
$$\eqalign{\left[ H_\alpha, H_\beta\right] &= 0 \cr
\left[ H_\alpha, E_{\pm\beta}\right] &= \pm K_{\alpha\beta} E_{\pm\beta} \cr
\left[ E_\alpha, E_{-\beta}\right] &= \delta_{\alpha\beta} H_\alpha\ \ .\cr}
\eqno(12)$$
Here $K_{\alpha\beta}$ is the (classical) Cartan matrix for the Lie algebra.
For $SU(N)$, $K$ is the $(N-1)\times(N-1)$ symmetric tridiagonal matrix
$$K_{\alpha\beta} = \left( \matrix{ \phantom{-}2 & -1 & \phantom{-}0 &
\phantom{-}\ldots
& \phantom{-}0 & \phantom{-}0 & \phantom{-}0  \cr
-1 & \phantom{-}2 & -1 & \phantom{-}\ldots & \phantom{-}0 & \phantom{-}0 &
\phantom{-}0
\cr
\phantom{-}0 & -1 & \phantom{-}2 & \phantom{-}\ldots & \phantom{-}0 &
\phantom{-}0 &
\phantom{-}0 \cr
\phantom{-}\vdots & \phantom{-}\vdots & \phantom{-}\vdots & &
\phantom{-}\vdots & \phantom{-}\vdots & \phantom{-}\vdots \cr
\phantom{-}0 &\phantom{-}0 & \phantom{-}0 & \phantom{-}\ldots & \phantom{-}2 &
-1 &
\phantom{-}0 \cr
\phantom{-}0 & \phantom{-}0 & \phantom{-} 0 & \phantom{-}\ldots &-1 &
\phantom{-}2 & -1
\cr
\phantom{-}0 & \phantom{-}0 & \phantom{-}0 & \phantom{-}\ldots & \phantom{-}0 &
-1 &
\phantom{-}2 \cr}\right)\ \ ,\eqno(13)$$
which is familiar from the discrete approximation to the second derivative in
numerical analysis.

With this choice (11) for the matter and gauge fields, the self-dual
Chern--Simons equations combine to yield the system of equations
$$\nabla^2 \ln\rho_\alpha = - {2\over\kappa} K_{\alpha\beta}\rho_\beta \qquad
(\alpha=1\ldots N-1)\ \ ,\eqno(14)$$
where $\rho_\alpha\equiv \left|\psi^\alpha\right|^2$.  This system is known as
the two-dimensional classical Toda equations.  For $SU(2)$, (14) becomes the
Liouville equation
$$\nabla^2 \ln\rho = - {4\over\kappa} \rho\ \ ,\eqno(15)$$
which Liouville showed to be integrable and indeed ``solvable''[ 13]
--- in the
sense that the general (real) solution could be expressed in terms of a single
arbitrary holomorphic function $f = f(x^-)$:
$$\rho = {\kappa\over 2}
\nabla^2 \ln \left( 1 + f(x^-) \bar{f} (x^+) \right) \ \ .\eqno(16)$$
Kostant [11], and Leznov and Saveliev [12]
showed that the two-dimensional Toda
equations (14) are integrable (with $K$ the Cartan matrix of any simple Lie
algebra), and that the solutions are intimately related to the representation
theory of the corresponding Lie algebra (see also Ref.~[15]).
The general (real) solutions for
$\rho_\alpha$ may be expressed in terms of $r$ arbitrary holomorphic
functions, where $r$ is the rank of the algebra.  Indeed, explicit formulas
may be given expressing the $\rho_\alpha$ as a matrix element of certain
path-ordered exponentials in the $\alpha^{\rm th}$ fundamental
representation [12].
 In Ref.~[3] the $SU(N)$ Toda solutions were expressed in an equivalent but
simpler form, more reminiscent of Liouville's solution (16) for the $SU(2)$
case:
$$\rho_\alpha\equiv {\kappa\over 2}\nabla^2 \ln\det \left( M^\dagger_\alpha
M_\alpha\right)\qquad (\alpha=1\ldots N-1)\ \ ,\eqno(17)$$
where $M_\alpha$ is the $N\times\alpha$ {\it rectangular\/} matrix
$$M_\alpha = \left( \matrix{ u & \partial_-u & {\partial_-}^2u & \ldots &
\partial^{\alpha-1}_- u\cr}\right)\ \ ,\eqno(18)$$
with $u$ being an $N$-component column vector
$$u = \left( \matrix{ 1 \cr f_1(x^-)\cr f_2(x^-)\cr\vdots \cr\
 f_{N-1}(x^-)\ \cr}
\right)\ \ ,\eqno(19)$$
involving $(N-1)$ arbitrary holomorphic functions $f_\alpha(x^-)$.  For $N=2$,
this trivially reduces to Liouville's solution (16).  We shall
discuss the general $SU(N)$  solution (17) -- (19) in more detail in
Section~6 in relation to the chiral model.

By extending the algebraic {\it Ans\"atz\/} (11b) for the matter fields (while
retaining (11a) for the gauge fields) to
$$\Psi = \sum\limits_{\alpha=
{{\scriptstyle{\rm positive}\atop \scriptstyle{\rm
simple}}\atop \scriptstyle{\rm roots}}}
\psi^\alpha E_\alpha + \psi^M E_{-M}\ \ ,\eqno(20)$$
where $E_{-M}$ is the step operator corresponding to minus the maximal root,
the self-dual Chern--Simons equations (9) and (10) combine into the {\it
affine\/} Toda equations
$$\nabla^2 \ln\rho_a = - {2\over\kappa}
\widetilde K_{ab} \rho_b\ \ ,\eqno(21)$$
where $\widetilde K_{ab}$ is the $(r+1)\times(r+1)$ {\it affine\/} Cartan
matrix.  For $SU(N)$, $\widetilde K_{ab}$ is the $N\times N$ symmetric matrix
$$\widetilde K_{ab}
= \left( \matrix{ \phantom{-}2 & -1 & \phantom{-}0 & \phantom{-}\ldots
& \phantom{-}0 & \phantom{-}0 & -1  \cr
-1 & \phantom{-}2 & -1 & \phantom{-}\ldots & \phantom{-}0 &
\phantom{-}0 & \phantom{-}0
\cr
\phantom{-}0 & -1 & \phantom{-}2 & \phantom{-}\ldots &
\phantom{-}0 & \phantom{-}0 &
\phantom{-}0 \cr
\phantom{-}\vdots & \phantom{-}\vdots & \phantom{-}\vdots & &
\phantom{-}\vdots & \phantom{-}\vdots & \phantom{-}\vdots \cr
\phantom{-}0 &\phantom{-}0 & \phantom{-}0 &
\phantom{-}\ldots & \phantom{-}2 & -1 &
\phantom{-}0 \cr
\phantom{-}0 & \phantom{-}0 & \phantom{-} 0 & \phantom{-}\ldots
&-1 & \phantom{-}2 & -1 \cr
-1 & \phantom{-}0 & \phantom{-}0 & \phantom{-}\ldots & \phantom{-}0 & -1 &
\phantom{-}2 \cr}\right)\ \ ,\eqno(22)$$

The {\it affine\/} Toda equations (23) are integrable, but it is not possible
to give convergent expressions for the general solutions in terms of a certain
number of arbitrary functions.
\goodbreak
\bigskip
\hangindent=21pt\hangafter=1
\noindent{\bf 4.\quad EQUIVALENCE BETWEEN SELF-DUAL CHERN--SIMONS\hfill\break
 EQUATIONS AND CHIRAL MODEL EQUATIONS}
\medskip
\nobreak
In Ref.~[3] it was shown that it is possible to make a gauge transformation
$g^{-1}$ (as in Eq.~(6)) which combines the self-dual Chern--Simons equations
into a
single matrix equation
$$\partial_- \chi = \left[\chi^\dagger,\chi\right]\ \ ,\eqno(23)$$
where
$$\chi = \sqrt{{2\over \kappa}} g \Psi \,g^{-1}\ \ .\eqno(24)$$
The existence of such a gauge transformation $g^{-1}$ follows from the
following zero-curvature relation.  To see this, define
$$\eqalignno{{\cal A}_+ &\equiv A_+ - \sqrt{{2\over\kappa}} \Psi
&(25\hbox{a}) \cr
{\cal A}_- &\equiv A_- + \sqrt{{2\over\kappa}} \Psi^\dagger\ \
.&(25\hbox{b})\cr}$$
Then the self-dual Chern--Simons equations imply that
$$
\partial_- {\cal A}_+ - \partial_+ {\cal A}_- + \left[ {\cal A}_-, {\cal
A}_+\right] = 0 \ \ ,\eqno(26)$$
which means that we can (locally) write ${\cal A}_\pm$ as
$${\cal A}_\pm = g^{-1} \partial_\pm g\ \ ,\eqno(27)$$
for some $g\in SU(N)$.  Then, defining $\chi$ as in (24), we see that (25) and
(27) imply that
$$D_- \Psi = \sqrt{{\kappa\over 2}}\ g^{-1} \left( \partial_- \chi - \left[
\chi^\dagger,\chi\right]\right) g\ \ ,\eqno(28)$$
and
$$\partial_- A_+ - \partial_+ A_- + \left[ A_-, A_+\right] - {2\over\kappa}
\left[ \Psi^\dagger,\Psi\right] = g^{-1} \left( \partial_-\chi + \partial_+
\chi^\dagger - 2\left[\chi^\dagger,\chi\right]\right) g\ \ .\eqno(29)$$
This shows that the self-dual Chern--Simons equations (9) and (10) are
equivalent to the single equation (23).

The equation (23) may now be rewritten as the {\it chiral model\/} equation:
$$\partial_+ \left( h^{-1} \partial_-h\right) + \partial_- \left( h^{-1}
\partial_+ h \right) = 0 \ \ ,\eqno(30)$$
where $h\in SU(N)$ is related to $\chi$ as:
$$\eqalignno{ h^{-1} \partial_+ h &= 2\chi &(31\hbox{a}) \cr
h^{-1} \partial_- h &= - 2\chi^\dagger\ \ .&(31\hbox{b}) \cr}$$
Note that if we define $J_+=2\chi$ and $J_-=-2\chi^\dagger$, then (23) (and
its conjugate) become
$$\eqalignno{\partial_+ J_- + \partial_- J_+ &= 0 \ \ ,&(32\hbox{a}) \cr
\partial_- J_+ - \partial_+ J_- + \left[ J_-, J_+\right] &= 0\ \
,&(32\hbox{b}) \cr}$$
which express the fact that $J$ has zero divergence and zero curl (in the
non-Abelian sense). For this reason, the chiral model
equations (30) are also known as the harmonic map equations.   Equation (32b)
shows that $J$ is a pure gauge, which justifies (31a,b) --- then the chiral
model equation (30) is simply the zero divergence equation
 (32a) expressed in terms of $h$.

We conclude this section by stressing that given {\it any\/} solution $h$ of
the
chiral model equation (30), the matrices $\chi$ and $\chi^\dagger$ defined in
(31a) automatically solve (23).  We thereby obtain a solution
$$\eqalignno{\Psi^{(0)} &= \sqrt{{\kappa\over 2}} \chi &(33\hbox{a}) \cr
A^{(0)}_+ &= \chi &(33\hbox{b}) \cr
A^{(0)}_- &= -\chi^\dagger\ \ ,&(33\hbox{c}) \cr}$$
of the self-dual Chern--Simons equations.

In order to compare these solutions with the Toda solutions discussed in the
previous section we note that, with the algebraic {\it Ans\"atze\/} (11) and
(20), the (Hermitian)
non-Abelian charge density $\rho=\left[\Psi,\Psi^\dagger\right]$ is
{\it diagonal\/} (it is also traceless and so it may be decomposed in terms of
the Cartan subalgebra elements).  In contrast, the solutions obtained from the
chiral model have charge density $\rho^{(0)} =
\kappa/2\left[\chi,\chi^\dagger\right]$ which need not be diagonal.  However,
$\rho^{(0)}$ {\it is\/} Hermitian and so it {\it can\/} be diagonalized by a
unitary matrix $g$.  This diagonalizing matrix is precisely the gauge
transformation relating $\Psi$ and $\chi$ in (24).  It is an algebraically
non-trivial task to construct explicitly this gauge transformation and obtain
the diagonal form of $\rho^{(0)}$ --- however, we shall present such an
explicit
diagonalization in Section~6, relating certain chiral model solutions to the
classical $SU(N)$ Toda solutions (17) -- (19).
\goodbreak
\bigskip
\noindent{\bf 5.\quad CLASSIFICATION OF SOLUTIONS}
\medskip
The main point of exhibiting the equivalence of the self-dual Chern--Simons
equations (9) -- (10) to the chiral model equation (30) is that {\it all\/}
solutions to the latter have been classified (subject to a finiteness
condition which has direct physical relevance in the Chern--Simons language).
Recall that this amounts to classifying {\it all\/} zero-energy static
solutions to the gauged nonlinear Schr\"odinger equation (1) with
Chern--Simons coupling (3).

In the two-dimensional Euclidean chiral model the ``classical action'' or
``energy functional'' is
$${\cal E}(h) = - {1\over 2} \int d^2x\, \tr \left( h^{-1} \partial_- h\,
h^{-1}
\partial_+ h\right)\ \ ,\eqno(34)$$
which is manifestly positive.  The classification (to be described in detail
below) of solutions to the chiral model equation (30) is achieved subject to
the finiteness condition
$${\cal E} (h) <\infty\ \ .\eqno(35)$$
Such a finiteness condition is appropriate in the $2+1$-dimensional
non-relativistic matter-Chern--Simons system ((1) and (3)) because
$$\eqalign{{\cal E}(h) &= 2\int d^2x\, \tr\left(\chi\chi^\dagger\right) \cr
&={4\over\kappa} \int d^2x\,\tr \left(\Psi\Psi^\dagger\right) \cr
&= {4\over\kappa} \int d^2x\,V^0 \cr
&\equiv {4\over\kappa}Q\ \ ,\cr}\eqno(36)$$
where $Q$ is the net gauge invariant charge, and
we have used the relations (31), (24) and (5).  Thus, the ``finite
energy'' condition (35) of the chiral model is precisely the ``finite charge''
condition of the Chern--Simons system.  As well as being physically
significant, the finiteness condition (35) is mathematically {\it crucial\/}
because the classification results of Uhlenbeck [19], Wood [23] and
Valli [20] are actually formulated for chiral model solutions on $S^2$,
rather than on ${\bf R}^2$.  However, when the ``charge'' (or ``action'') in
(36) is finite, $h$ extends to the conformal compactification ${\bf
R}^2\cup\{\infty\} = S^2$ of ${\bf R}^2$, and so the classification of finite
charge solutions on ${\bf R}^2$ is equivalent to the classification of
solutions on $S^2$ [22,17].    This fortuitous correspondence permits
us to take over directly the following results from the mathematical
literature regarding the classification of solutions to the chiral model
equations.
\medskip
{\noindent\narrower{\bf Theorem} (K.~Uhlenbeck [19];
see also J.~C.~Wood [23]): {\sl Every finite
action solution $h$ of the $SU(N)$ chiral model equation (30) may be
uniquely factorized as a product
$$h = \pm h_0 \prod\limits^m_{i=1} \left( 2p_i-1\right) \eqno(37)$$
where:
\item{a)}$h_0\in SU(N)$ is constant;
\smallskip
\item{b)}each $p_i$ is a Hermitian projector ($p^\dagger_i = p_i$ and $p^2_i =
p_i$);
\smallskip
\item{c)}defining $h_j = h_0 \prod\limits^j_{i=1} (2p_i-1)$, the following
{\it linear\/} relations must hold:
$$\eqalign{(1-p_i) \left( \partial_+ + {1\over 2} h^{-1}_{i-1} \partial_+
h_{i-1}\right) p_i &= 0\ \ ,\cr
\left( 1-p_i\right) h^{-1}_{i-1} \partial_- h_{i-1} p_i &= 0\ \ ;\cr}$$
\item{d)}$m\le N-1$.\smallskip}}
\medskip
\noindent Each factor $(2p_i-1)$ is referred to as a ``uniton'' factor.

The $\pm$ sign in (37) has been inserted to allow for the fact that Uhlenbeck
and Wood considered the case of $U(N)$ rather than $SU(N)$.  However, since
$\det(2p_i-1)=(-1)^{N-d_i}$ if $p_i$ is an $N\times N$ Hermitian projector
onto a $d_i$-dimensional subspace, we see that we can obtain a solution $h\in
SU(N)$ simply by choosing $h_0\in SU(N)$ and the appropriate overall $\pm$
sign.

Note that Uhlenbeck's theorem tells us that for $SU(2)$ {\it all\/} finite
action solutions of the chiral model have the form
$$h = - h_0(2p-1) \eqno(38)$$
where $p$ is a holomorphic projector
$$\left( 1 - p\right) \partial_+ p = 0\ \ .\eqno(39)$$
Since $p^2 = p$, condition (39) is equivalent to $\partial_+ p\,p=0$.  All
such holomorphic projectors may be written as the projection matrix
$$p = M \left( M^\dagger M\right)^{-1} M^\dagger \eqno(40)$$
where $M=M(x^-)$ is any rectangular matrix depending only on the $x^-$
variable (so $M^\dagger=M^\dagger(x^+)$ is a function only of $x^+$).  It is
straightforward to verify that such a projector satisfies (39).  In order to
obtain an $h$ in the defining representation of $SU(2)$ we choose
$$M = \left( \matrix{ 1\cr\noalign{\vskip 0.2cm} f(x^-)\cr}\right) \eqno(41)$$
so that
$$p = {1\over 1+f\bar{f}} \left( \matrix{ 1 & \bar{f}\cr\noalign{\vskip
0.2cm}f&f\bar{f}\cr}\right)\ \ .\eqno(42)$$
(Note that in general one should consider projection onto the space spanned by
$M = \left( \matrix{ f_1(x^-)\cr f_2(x^-)\cr}\right)$, but since only the {\it
direction\/} is important for $p$, this reduces to either (41) or to
$\left(\matrix{f(x^-)\cr1\cr}\right)$.  The formulas in the latter case are
analogous, and we shall in fact see that in the final result (48) there is no
distinction --- a {\it single\/} function $f(x^-)$ suffices to determine the
digaonal form of $\left[\chi,\chi^\dagger\right]$.)

The corresponding solution $\chi$ of the self-dual Chern--Simons equation is
given by (see (31))
$$\eqalign{\chi &= {1\over 2} h^{-1}\partial_+ h \cr
&= {1\over 2} (2p-1) \cdot 2\partial_+ p\cr
&= 2p\partial_+p - \partial_+p\cr
&= \partial_+ p\cr
&= {f\partial_+\bar{f}\over \left( 1 + f\bar{f}\right)^2} \left( \matrix{ -1 &
{1\over f}\cr\noalign{\vskip 0.2cm} -f & 1 \cr} \right)\ \ .\cr}\eqno(43)$$
The corresponding matter density is
$$\left[ \chi,\chi^\dagger\right] = {\partial_+ \bar{f} \partial_- f\over
\left(
1 + f\bar{f}\right)^3} \left( \matrix{ 1 - f\bar{f} &
2\bar{f}\cr\noalign{\vskip 0.2cm} 2f & -1+f\bar{f} \cr}\right)\ \ .\eqno(44)$$
This may be diagonalized using the unitary matrix
$$g = {1\over \sqrt{1 + f\bar{f}}} \left( \matrix{ 1 &
-\bar{f}\cr\noalign{\vskip 0.2cm} f & \phantom{-}1\cr}\right)\eqno(45)$$
which also diagonalizes $p$:
$$\eqalignno{g^{-1} pg &= \left( \matrix{ 1 & 0 \cr\noalign{\vskip 0.2cm} 0 &
0 \cr}\right) &(46) \cr\noalign{\vskip 0.2cm}
g^{-1} \chi g &= {\partial_+ \bar{f}\over \left( 1 + f\bar{f}\right)} \left(
\matrix{ 0 & 1 \cr\noalign{\vskip 0.2cm} 0 & 0 \cr}\right)
 &(47)\cr\noalign{\vskip 0.2cm}
g^{-1} \left[ \chi,\chi^\dagger\right]g &= {\partial_+ \bar{f} \partial_-
f\over \left( 1 + f\bar{f}\right)^2} \left( \matrix{ 1 & \phantom{-}0
\cr\noalign{\vskip 0.2cm} 0 & -1 \cr}\right) \cr
&= \partial_+ \partial_- \ln \left( 1 + f\bar{f}\right) \left( \matrix{ 1 &
\phantom{-}0\cr\noalign{\vskip 0.2cm} 0 & -1\cr}\right) \ \ .\cr
&= \partial_+\partial_- \ln \det \left( M^\dagger M\right) \left( \matrix{ 1 &
\phantom{-}0\cr\noalign{\vskip 0.2cm} 0 & -1\cr}\right)\ \ .&(48)\cr}$$
But this is precisely the classical
$SU(2)$ Toda solution (16).  Thus we see that the
$SU(2)$ element $g$ in (45) is the gauge transformation which converts
between the classical
Toda solution obtained using the {\it Ans\"atz\/} (11) (compare
with (47)) and Uhlenbeck's chiral model solution (38) with $p$ as in (42).
Interestingly, Uhlenbeck's result tells us that this classical
Toda solution is the {\it
only\/} one with finite charge for $SU(2)$  (note that the charge is gauge
invariant).  In particular, as was argued in Ref.~[3], there is no finite
charge solution for the $SU(2)$ {\it affine\/} Toda equation arising from the
algebraic {\it Ans\"atz\/} in (20) (this is consistent with the index theory
analysis of Ref.~[24]).    In the next section we will construct the
analogous gauge transformation relating a particular solution of the $SU(N)$
chiral model equations in Uhlenbeck's factorized form (37), with the {\it
explicit\/} $SU(N)$ Toda solutions (17) -- (19).

It becomes significantly more involved to describe systematically
all possible uniton
factorizations of solutions to the $SU(N)$ chiral model equations for $N\ge
3$.  Wood [23] has given an explicit construction and parametrization of {\it
all\/} $SU(N)$ solutions in terms of sequences of Grassmannian factors. The
parameterization involves an algorithm which uses only algebraic, derivative
and integral transform operations.  The $N=3$ and $N=4$ [17,23] cases have
been studied in great detail.  In the next section we present the uniton
factorization of a general class of harmonic maps for {\it any\/} $N$.

To conclude this section we quote a result due to
Valli:
\medskip
{\noindent\narrower{\bf Theorem} (G.~Valli [20]): {\sl Let $h$ be
a solution of the chiral
model equation (30).  Then the action
$$-{1\over 2}\int d^2x\, \tr \left( h^{-1} \partial_- h\,h^{-1} \partial_+ h
\right) $$
is quantized in {\it integral\/} multiples of $8\pi$.\smallskip}}
\medskip
\noindent Give the relation (36) between the chiral model action and the
Chern--Simons model charge $Q$, we obtain, as a corollary of Valli's theorem,
the result that the charge is quantized in integral multiples of $2\pi\kappa$.
This fact had been independently noted (see Appendix of Ref.~[3]) for the
classical $SU(N)$ Toda solutions.
\goodbreak
\bigskip
\noindent{\bf 6.\quad TODA SOLUTIONS AND THE CHIRAL MODEL}
\medskip
\nobreak
In this section we generalize the single uniton $SU(2)$ solution discussed in
(38) -- (48) to a multi-uniton $SU(N)$ solution.  This solution to the chiral
model equation (30) has the remarkable property that it is gauge equivalent to
the $SU(N)$ Toda solutions in (17) -- (19).  Let us first state the result,
and then prove its validity.
\medskip
{\noindent\narrower {\bf Main Result:}\quad \sl The following matrix
$$h =(-1)^{{1\over 2} N(N+1)}
 \prod\limits^{N-1}_{\alpha=1} \left( 2p_\alpha-1\right)\ \ ,\eqno(49)$$
where $p_\alpha$ is the Hermitian holomorphic projector $p_\alpha = M_\alpha
\left( M^\dagger_\alpha M_\alpha\right)^{-1} M^\dagger_\alpha$ for the matrix
$M_\alpha$ in (18), is a solution of the chiral model equation $\partial_+
\left( h^{-1} \partial_- h\right) + \partial_-\left( h^{-1} \partial_+
h\right) = 0$.
Furthermore, with $\chi$ and $\chi^\dagger$ related to $h$ as in (31), ({\it
i.e.\/} $\chi\equiv {1\over 2} h^{-1} \partial_+ h$, $\chi^\dagger\equiv -
{1\over 2} h^{-1} \partial_- h$) there
exists a unitary transformation $g$ which diagonalizes the charge density
matrix $\left[ \chi,\chi^\dagger\right]$ so that
$$g^{-1} \left[\chi,\chi^\dagger\right]g = \sum\limits^{N-1}_{\alpha=1}
\partial_+\partial_- \ln\det \left( M^\dagger_\alpha M_\alpha\right) H_\alpha\
\ ,\eqno(50)$$
where $H_\alpha$ are the CSA generators of $SU(N)$ in the Chevalley basis for
the defining representation.  Recalling (11) and (17) we recognize this
diagonalized form (50)
as the $SU(N)$ Toda solution to the self-dual Chern--Simons
equations.  \smallskip}
\medskip

\noindent {\bf Proof:}\quad The matrix $h$ in (49) is clearly unitary, and the
$(-1)^{{1\over 2}N(N+1)}$ factor ensures that $h\in SU(N)$.  The
matrix $g$ appearing in (50) (the $SU(N)$ matrix which provides the
gauge transformation between the chiral model and Toda solutions) is simply
the unitary matrix
$$g = \left( e_1\ e_2\ \ldots\ e_N\right)\eqno(51)$$
which simultaneously diagonalizes all the $p_\alpha$.
Thus the column vectors $\left\{ e_\alpha\right\}$ are just the orthonormal
basis elements
constructed by the Gramm--Schmidt process beginning with the column vectors
$u$, $\partial_-u$, $\partial^2_-u,\ldots \partial^{N-1}_-u$ (which are
assumed to be linearly independent).  Since the vectors $e_1,\ldots e_\alpha$
span the same space as the vectors $u$,
$\partial_-u,\ldots,\partial^{\alpha-1}_-u$ it is clear that
$$p_\alpha = \widetilde M_\alpha \left( \widetilde M^\dagger_\alpha
\widetilde M_\alpha\right)^{-1} \widetilde M^\dagger_\alpha\eqno(52)$$
where $\widetilde M_\alpha=\left(e_1\ e_2\ \ldots\  e_\alpha\right)$.
And since the
$e_\alpha$'s are orthonormal $\left(e^\dagger_\alpha
e_\beta=\delta_{\alpha\beta}\right)$
we find a simple expression for $p_\alpha$:
$$p_\alpha = \sum\limits^\alpha_{\beta=1} e_\beta e^\dagger_\beta\ \
.\eqno(53)$$
Note however that the column vectors $e_\alpha$ depend on {\it both\/} $x^-$
and
$x^+$ (unlike the column vectors $\partial^\alpha_-u$ which only depend on
$x^-$) --- this dependence enters through the Gramm--Schmidt procedure:
$$e_\alpha = {\left( 1 - p_{\alpha-1}\right) \partial^{\alpha-1}_- u\over
\sqrt{ \partial^{\alpha-1}_+ u^\dagger \left( 1 - p_{\alpha-1}\right)
\partial^{\alpha-1}_- u}}\ \ ,\qquad (\alpha=1\ldots N) \eqno(54)$$
where $p_0\equiv0$.  The unitary matrix $g$ in (51) diagonalizes {\it each\/}
$p_\alpha$ projection matrix.  In fact, due to the orthonormality of the
columns it is easy to see that
$$g^{-1} p_\alpha g =
\left( \matrix{
1\cr
&1 & & & & & 0 \cr
& & \ddots\cr
& & & 1\cr
& & & & 0\cr
& & & & & \ddots \cr
0 & & & & & & 0 \cr}\right)\eqno(55)$$
where the first $\alpha$ entries on the diagonal are 1, all other entries
being 0.  (This is hardly surprising, as $p_\alpha$ is a projector onto the
$\alpha$-dimensional subspace spanned by $e_1, \ldots e_\alpha$).

To verify that $h$ in (49) does indeed solve the chiral model equation (30) we
first show that
$$h^{-1} \partial_+ h = 2\sum\limits^{N-1}_{\alpha=1} \partial_+ p_\alpha\ \
.\eqno(56)$$
Then, since $h$ is unitary and each $p_\alpha$ is Hermitian, we deduce that
$$h^{-1} \partial_- h = - 2 \sum^{N-1}_{\alpha=1} \partial_- p_\alpha\ \
,\eqno(57)$$
from which the chiral model relation $\partial_+
\left(h^{-1} \partial_- h\right) + \partial_- \left( h^{-1} \partial_+h\right)
= 0$ follows immediately.

To prove (56) we first note that each $p_\alpha$ is a holomorphic projector:
{\it i.e.\/} $p_\alpha\partial_+ p_\alpha=\partial_+ p_\alpha$
for each $\alpha=1\ldots N-1$.
Therefore, from (49) we have
$$\eqalign{h^{-1}\partial_+ h &= \sum^{N-1}_{\beta=1} \left\{ \left(
2p_{N-1}-1\right)\ldots \left( p_{\beta+1}-1\right)
\left( 2p_\beta-1\right) 2 \partial_+ p_\beta \left(
2p_{\beta+1}-1\right) \ldots \left( 2p_{N-1}-1\right)\right\} \cr
&= \sum^{N-1}_{\beta=1} \left\{ \left( 2p_{N-1}-1\right)\ldots \left(
2p_{\beta+1}-1\right) 2\partial_+ p_\beta \left( 2p_{\beta+1}-1\right) \ldots
\left( 2p_{N-1}-1\right)\right\} \ \ .\cr} \eqno(58)$$
The result (56) follows now if we can prove that
$$\left[ \partial_+ p_\alpha,p_\beta\right]=0\qquad \forall \alpha<\beta\ \
.\eqno(59)$$
{}From (55) we have that $\partial_+ \left( g^{-1} p_\alpha g\right) = 0$, or
equivalently:
$$g^{-1}\left(\partial_+ p_\alpha\right) g
= \left[ g^{-1}\partial_+ g,g^{-1} p_\alpha g\right]\ \ ,\eqno(60)$$
and we note the particularly simple form of $g^{-1} p_\alpha g$ as in (55). Now
$g^{-1}\partial_+ g$ is an $N\times N$ matrix whose $\left(
\alpha\beta\right)^{\rm th}$ entry is $e^\dagger_\alpha \partial_+ e_\beta$.
{}From the Gramm--Schmidt procedure (54) it is clear that $\partial_+ e_\beta$
is a linear combination of $e_1\ldots e_\beta$, and so, by the orthonormality
of the basis,
$$e^\dagger_\alpha \partial_+ e_\beta = 0 \qquad \forall\ \  \alpha>\beta\ \
.\eqno(61)$$
Similarly, since $\partial_+ \left( e^\dagger_\alpha e_\beta\right) =
\partial_+ \left( \delta_{\alpha\beta}\right) = 0$,
$$\eqalign{e^\dagger_\alpha \partial_+ e_\beta &=-\partial_+ e^\dagger_\alpha
e_\beta \cr
&= - \left( e^\dagger_\beta \partial_- e_\alpha\right)^\dagger\cr
&= 0 \qquad \forall\ \  \beta>\alpha+1\ \ ,\cr}\eqno(62)$$
where in the last step we used the fact that
$\partial_-e_\alpha$ is a linear combination of $e_1,\ldots,
e_{\alpha+1}$.  Thus, the matrix $g^{-1}\partial_+ g$ has the following
simple form, with non-zero entries only on and immediately above the diagonal:
$$g^{-1}\partial_+ g = \left(\matrix{
e^\dagger_1\partial_+ e_1 & e^\dagger_1 \partial_+ e_2 & & 0\cr
\ddots & \ddots \cr
& e^\dagger_\alpha \partial_+ e_\alpha & e^\dagger_\alpha\partial_+
e_{\alpha+1} \cr
& \ddots & \ddots \cr
0&&&e^\dagger_{N-1}\partial_+ e_N\cr
&&&e^\dagger_N \partial_+ e_N\cr}\right)\ \ . \eqno(63)$$
Equations (60) and (55) then imply $g^{-1}\left(\partial_+ p_\alpha\right) g$
has only {\it one\/} non-zero entry, in the $\left(
\alpha(\alpha+1)\right)^{\rm th}$ place:
$$g^{-1}\left(\partial_+ p_\alpha\right) g = \left(\matrix{
0\cr
&\ddots\cr
&&0 & e^\dagger_\alpha\partial_+ e_{\alpha+1}\cr
 & & 0 & 0 \cr
&&&&\ddots\cr
&&&&0\cr}\right)\ \ .\eqno(64)$$
It follows that $\left[ g^{-1}\left(\partial_+ p_\alpha\right) g,\;g^{-1}
p_\beta g\right]=0$ for $\alpha<\beta$, which is just the statement (59).

This completes the proof that
the matrix $h$ in (49) does indeed satisfy the chiral model equation.  We now
proceed to prove the diagonalization formula (50).  Observe that
$$\eqalign{g^{-1} \chi g &= {1\over 2} g^{-1} \left( h^{-1} \partial_+
h\right) g \cr
&= \sum\limits^{N-1}_{\alpha=1} g^{-1} \left(\partial_+p_\alpha\right) g\cr
&= \sum^{N-1}_{\alpha=1} \left( e^\dagger_\alpha \partial_+
e_{\alpha+1}\right) E_\alpha \ \ ,\cr}\eqno(65)$$
where $E_\alpha$ are the $SU(N)$ positive simple root step operators in the
defining representation: $\left( E_\alpha\right)_{ab} = \delta_{\alpha
a}\delta_{\alpha+1,b}$.  It should be noted
that this is of the same algebraic {\it Ans\"atz\/} form as the $SU(N)$ {\it
classical\/} Toda {\it Ans\"atz\/} for the matter field $\Psi$ in (11b) for
the Chern--Simons model.  To make the comparison complete we note that
$$e^\dagger_\alpha \partial_+ e_{\alpha+1} = \sqrt{\partial_+ \partial_-
\ln\det \left( M^\dagger_\alpha M_\alpha\right)}\ \ .\eqno(66)$$
This may be verified directly from the Gramm--Schmidt projections (54) or by
noting that
$$\eqalign{
\partial_+\partial_-\ln\det \left( M^\dagger_\alpha M_\alpha\right) &=
\partial_+ \partial_- \tr\ln \left( M^\dagger_\alpha M_\alpha \right) \cr
&= \partial_+ \tr \left( \left( M^\dagger_\alpha M_\alpha\right)^{-1}
M^\dagger_\alpha \partial_- M_\alpha\right) \cr
&= \tr \left(\left( 1 - p_\alpha\right) \partial_- M_\alpha
\left(M^\dagger_\alpha M_\alpha\right)^{-1} \partial_+ M^\dagger_\alpha \left(
1 - p_\alpha\right)\right) \cr
&= \tr \left( \partial_- p_\alpha \partial_+ p_\alpha\right)\cr
&= \left| e^\dagger_\alpha \partial_+ e_{\alpha+1}\right|^2\ \
,\cr}\eqno(67)$$
where in the last step we have used (64) (and its conjugate).

It now follows immediately that the diagonalized charge density is
$$g^{-1} \left[ \chi, \chi^\dagger\right] g = \sum\limits^{N-1}_{\alpha=1}
\partial_+\partial_- \ln\det \left( M^\dagger_\alpha M_\alpha\right) H_\alpha\
\
,\eqno(68)$$
as claimed in (50).  This should be compared with the $SU(N)$ Toda solution
(17) (see (48) for the explicit $SU(2)$ case).
Finally, the net Abelian charge is
$$\eqalign{Q&={\kappa\over 2} \int d^2x \tr \left(\chi\chi^\dagger\right) \cr
&= {\kappa\over 2}
\sum^{N-1}_{\alpha=1} \int d^2x\,\partial_+ \partial_- \ln \det \left(
M^\dagger_\alpha M_\alpha\right) \cr
&= \sum^{N-1}_{\alpha=1} \int d^2x\, \rho_\alpha\cr}\eqno(69)$$
with the $\rho_\alpha$ being the non-Abelian charge densities of the Toda
solution (17).
\goodbreak
\bigskip
\noindent{\bf 7.\quad CONCLUDING COMMENTS}
\medskip
In summary, we have shown that the static, self-dual zero-energy solutions of
the $2+1$-dimensional gauged nonlinear Schr\"odinger equation (1), with
Chern--Simons coupling (3), may be classified in terms of the Uhlenbeck--Wood
classification of solutions to the chiral model (or harmonic map) equations
(30).  The gauge invariant charge is quantized in integral multiples of
$2\pi\kappa$.  We have also found the explicit uniton factorization of a
general class of harmonic maps into $SU(N)$, this class being distinguished by
the fact that the corresponding matter density matrix, when diagonalized,
satisfies the classical $SU(N)$ Toda equations.

For $SU(2)$, the Toda-type solutions exhaust {\it all\/} finite charge
solutions, while for the $SU(N)$ $N\ge 3$ systems this is {\it not\/} the case
--- there are harmonic map solutions which are not of Toda-type.  This may be
seen already for $SU(3)$ using the general solutions in Refs.~[23,17];
however, the non-Toda solutions for $SU(3)$ are somewhat awkward to present
explicitly.  The simplest non-Toda harmonic map solution arises in the $SU(4)$
model when we choose a one-uniton solution $h=2p-1$ with $p$ being a
holomorphic projector onto a two-dimensional holomorphic subspace.
(Note that
for $SU(3)$ we can only project onto a one-dimensional subspace, a
two-dimensional subspace simply being the orthogonal complement of some
one-dimensional subspace --- thus $2p-1$ just changes sign.) Thus $p$
may be written as $p=M\left(M^\dagger M\right)^{-1}M^\dagger$
where $M$ is a $4\times 2$
holomorphic matrix whose column space may be specified by six arbitrary
holomorphic functions.  Then the (gauge invariant!) charge density $V^0$ is
given by $V^0 = \partial_+\partial_-\ln\det \left(M^\dagger M\right)$ also
depending on six functions.  However, the $SU(4)$ Toda solution has $V^0$ as
in (69) depending on only three arbitrary holomorphic functions (recall
$M_\alpha$ is given by (18) -- (19)).  This Toda
solution is therefore less general.

The relationship between the chiral model and Toda equations is especially
intriguing when viewed in light of the importance of the two-dimensional
self-duality equations to the theory of integrable partial differential
equations in two dimensions [21].  Ward [22] has also shown that the harmonic
map equations may be understood in the setting of algebraic geometry, adapting
results from ``twistor'' constructions of monopoles and Yang--Mills
instantons.  He had concentrated on the $SU(2)$ case --- it would be
interesting to see if similar constructions for $SU(N)$ of the Toda-type
harmonic maps presented here leads to a deeper understanding of the Toda
and/or chiral model equations.

Yet another mysterious correspondence arises from the work of Forg\'acs {\it
et al.\/} [5] and O'Raifeartaigh {\it et al.\/} [16],
who have shown that the Toda systems arise as certain special
cases not of the chiral model but of the Wess--Zumino--Novikov--Witten model.
It is unclear what is the connection between their results and those presented
here.

Finally, it would be interesting to consider the generalization of these
results to Lie algebras other than $SU(N)$.  For the compact simple algebras,
the special Toda solutions to the self-dual Chern--Simons equations are known
[11,12].  Can one find the transformation $g^{-1}$ which transforms these
solutions to chiral model solutions, thereby obtaining explicit harmonic maps
into the corresponding Lie group?
\goodbreak
\bigskip
\centerline{\bf ACKNOWLEDGEMENTS}
\medskip
I thank Professor R.~Jackiw for discussions and special thanks to Professor
R.~Ward for helpful correspondence.
\vfill
\eject
\centerline{\bf REFERENCES}
\medskip
\item{1.}A. Bilal and J.-L. Gervais, ``Extended $c=\infty$ Conformal Systems
from Classical Toda Field Theories,'' {\it Nucl. Phys.\/} {\bf B314}, 646--686
(1989).
\medskip
\item{2.}S. Donaldson, ``Twisted Harmonic Maps and the Self-Duality
Equations,'' {\it Proc. London Math. Soc.\/} {\bf 55}, 127--131 (1987).
\medskip
\item{3.}G. Dunne, R. Jackiw, S.-Y. Pi and C. Trugenberger, ``Self-Dual
Chern--Simons Solitons and Two-Dimensional Nonlinear Equations,'' {\it Phys.
Rev.\/} {\bf D43}, 1332--1345 (1991).
\medskip
\item{4.}G. Dunne, ``Self-Duality and Chern--Simons Theories,'' in {\it
Proceedings of the XX$^{\rm th}$ International Conference on Differential
Geometric Methods in Theoretical Physics\/}, S. Catto and A. Rocha, eds.
 (Baruch College, New York, June, 1991).
\medskip
\item{5.}P. Forg\'acs, A. Wipf, J. Balog, F. Feh\'er and L. O'Raifeartaigh,
``Liouville and Toda Theories as Conformally Reduced WZNW Theories,'' {\it
Phys. Lett.\/} {\bf B227}, 214--220 (1989); ``Toda Theory and $W$ Algebra from
a Gauged WZNW Point of View'', {\it Ann. Phys.\/} (NY) {\bf 203}, 76--136
(1990).
\medskip
\item{6.}B. Grossman, ``Hierarchy of Soliton Solutions to the Gauged Nonlinear
Schr\"odinger Equation on the Plane,'' {\it Phys. Rev. Lett.\/} {\bf 65},
3230--3232 (1990).
\medskip
\item{7.}N. Ganoulis, P. Goddard and D. Olive, ``Self-Dual Monopoles and Toda
Molecules,'' {\it Nucl. Phys.\/} {\bf B205}, 601--636 (1982).
\medskip
\item{8.}N. Hitchin, ``The Self-Duality Equations on a Riemann Surface,'' {\it
Proc. London Math. Soc.\/} {\bf 55}, 59--126 (1987).
\medskip
\item{9.}R. Jackiw and S.-Y. Pi, ``Soliton Solutions to the Gauged Non-Linear
Schr\"odinger Equation on the Plane,'' {\it Phys. Rev. Lett.\/} {\bf 64},
2969--2972 (1990); ``Classical and Quantum Non-Relativistic Chern--Simons
Theory,'' {\it Phys. Rev.\/} {\bf D42}, 3500--3513 (1990).
\medskip
\item{10.}See {\it e.g.\/} J. Humphreys, {\it Introduction to Lie Algebras and
Representation Theory\/} (Springer-Verlag, Berlin, 1990).
\medskip
\item{11.}B. Kostant, ``The Solution to a Generalized Toda Lattice and
Representation Theory,'' {\it Adv. Math.\/} {\bf 34}, 195--338 (1979).
\medskip
\item{12.}A. Leznov and M. Saveliev, ``Representation of Zero Curvature for
the System of Nonlinear Partial Differential Equations $x_{\alpha,z\bar{z}}
=\exp (kx)_\alpha$ and its Integrability,'' {\it Lett. Math. Phys.\/} {\bf 3},
389--494 (1979); ``Representation Theory and Integration
of Nonlinear Spherically Symmetric Equations of Gauge Theories,'' {\it Commun.
Math. Phys.\/} {\bf 74}, 111--118 (1980).
\medskip
\item{13.}J. Liouville, ``Sur l'\'equation aux diff\'erences partielles
${d^2\over du\,dv} \log\lambda\pm {\lambda\over 2a^2} = 0$,'' {\it Journ.
Math. Pures, Appl.\/} {\bf 18}, 71--72 (1853).
\medskip
\item{14.}P. Mansfield, ``Solution of Toda Systems,'' {\it Nucl. Phys.\/}
{\bf B208}, 277--300 (1982).
\medskip
\item{15.}A. Mikhailov, M. Olshanetsky and A. Perelomov, ``Two-Dimensional
Generalized Toda Lattice,'' {\it Commun. Math. Phys.\/} {\bf 79}, 473--488
(1981).
\medskip
\item{16.}L. O'Raifeartaigh, P. Ruelle, I. Tsutsui and A. Wipf, ``$W$-Algebras
for Generalized Toda Theories,'' {\it Commun. Math. Phys.\/} {\bf 143},
333--354 (1992).
\medskip
\item{17.}B. Piette and W. Zakrzewski, ``General Solutions of the $U(3)$ and
$U(4)$ Chiral Sigma Models in Two Dimensions,'' {\it Nucl. Phys.\/} {\bf
B300}, 207--222 (1988);
 ``Some Classes of General Solutions of
the $U(N)$ Chiral $\sigma$ Models in Two Dimensions,.'' {\it Journ. Math.
Phys.\/} {\bf 30}, 2233--2237 (1989).
\medskip
\item{18.}M. Toda, ``Studies of a Nonlinear Lattice,'' {\it Phys. Rep.\/} {\bf
8}, 1--125 (1975).
\medskip
\item{19.}K. Uhlenbeck, ``Harmonic Maps into Lie Groups (Classical Solutions
of the Chiral Model),'' {\it Jour. Diff. Geom.\/} {\bf 30}, 1--50 (1989).
\medskip
\item{20.}G. Valli, ``On the Energy Spectrum of Harmonic Two-Spheres in
Unitary Groups,'' {\it Topology\/} {\bf 27}, 129--136 (1988).
\medskip
\item{21.}R. S. Ward, ``Integrable and Solvable Systems and Relations Among
Them,'' {\it Phil. Trans. Roy. Soc. London\/} {\bf A315}, 451--457 (1985).
\medskip
\item{22.}R. S. Ward, ``Classical Solutions of the Chiral Model, Unitons and
Holomorphic Vector Bundles,'' {\it Comm. Math. Phys.\/} {\bf 128}, 319--332
(1990).
\medskip
\item{23.}J. C. Wood, ``Explicit Construction and Parametrization of Harmonic
Two-Spheres in the Unitary Group,'' {\it Proc. London Math. Soc.\/} {\bf 58},
608--624 (1989).
\medskip
\item{24.}Y. Yoon, ``Zero Modes of the Non-Relativistic Self-Dual
Chern--Simons Vortices on the Toda Backgrounds,'' {\it Ann. Phys.\/} (NY) {\bf
211}, 316--333 (1991).
\par
\vfill
\end